\documentclass[12pt]{iopart}

\usepackage{iopams}
\usepackage{hyperref}
\begin{document}

\title[Red light for Anderson localization]{Red light for Anderson localization$^{\dagger}$}
\footnotetext{This is an author-created, un-copyedited version of an article published in \NJP {\bf 18} (2016) 021001. IOP Publishing Ltd is not responsible for any errors or omissions in this version of the manuscript or any version derived from it. The Version of Record is available online at \url{http://dx.doi.org/10.1088/1367-2630/18/2/021001}.}

\author{S E Skipetrov$^{1,2}$ \& J H Page$^3$}
\address{$^1$Universit\'{e} Grenoble Alpes, LPMMC, F-38000 Grenoble, France}
\address{$^2$CNRS, LPMMC, F-38000 Grenoble, France}
\address{$^3$Department of Physics and Astronomy, University of Manitoba, Winnipeg, Manitoba R3T 2N2, Canada}
\ead{sergey.skipetrov@lpmmc.cnrs.fr, john.page@umanitoba.ca}

\begin{abstract}
During the last 30 years, the search for Anderson localization of light in three-dimensional (3D) disordered samples yielded a number of experimental observations that were first considered successful, then disputed by opponents, and later refuted by their authors. This includes recent results for light in TiO$_2$ powders that T. Sperling {\it et al} now show to be due to fluorescence and not to Anderson localization (\NJP {\bf 18} (2016) 013039). The difficulty of observing Anderson localization of light in 3D may be due to a number of factors: insufficient optical contrast between the components of the disordered material, near-field effects, etc. The way to overcome these difficulties may consist in using partially ordered materials, complex structured scatterers, or clouds of cold atoms in magnetic fields.
\end{abstract}

%
%
%
%
%

Anderson localization is a wave interference phenomenon leading to a breakdown of wave propagation in strongly disordered media \cite{anderson58}. It 
may occur for all types of waves: sound, microwaves, light or ``Schr\"{o}dinger'' waves corresponding to the wave functions of quantum particles---electrons or atoms---at low temperatures \cite{lagendijk09}. The possibility of stopping light with disorder was suggested three decades ago by Sajeev John \cite{john84} and the inventor of the localization phenomenon himself \cite{anderson85}, and was realized in a number of beautiful experiments by several groups \cite{segev13}. The undeniably successful experiments were all, however, performed in low-dimensional (1D or 2D) systems, whereas the most interesting case of 3D disorder has been a subject of controversial experiments and heated discussions since the first claim of observation of Anderson localization of light in GaAs powders by D.S. Wiersma {\it et al} in 1997 \cite{wiersma97}. The main objection 
was that the experimental signatures of localization reported in ref.\ \cite{wiersma97} could be equally well attributed to a weak absorption of light in the disordered sample \cite{scheffold99}. To refute this criticism and to separate the impacts of absorption and localization, time-of-flight experiments were performed on semiconductor samples similar to those of ref.\ \cite{wiersma97}, but no signature of Anderson localization was found \cite{johnson03,vanderbeek12}.

A new series of experiments exhibiting deviations from normal diffusive transport, which were interpreted as signatures of 3D Anderson localization of light, has been carried out since 2006 in the group headed by G. Maret \cite{storzer06,sperling13}. These experiments used pressed TiO$_2$ powders---samples similar in structure to simple white paints and thus directly inspired by the initial proposal by P.W. Anderson \cite{anderson85}. Pulsed light sources and time-dependent measurements allowed for separation of localization and absorption effects \cite{storzer06} or even, in transverse confinement experiments designed by analogy with the ultrasonic case \cite{hu08}, to the full independence of the measured quantities from absorption \cite{sperling13}. However, the optical contrast between scattering particles having refractive index $n \simeq 2.8$ and the surrounding medium (air) is lower than in semiconductor samples used in refs.\ \cite{wiersma97,johnson03,vanderbeek12} ($n \simeq 3.2$--3.6) making it difficult to understand that Anderson localization takes place in TiO$_2$ \cite{storzer06,sperling13} but apparently not in semiconductors \cite{johnson03,vanderbeek12}. Doubts were also expressed about the interpretation of the data, raising concerns about the role of inelastic scattering \cite{scheffold13}. Now, in a carefully executed and comprehensive experimental study, T. Sperling {\em et al} show that the results of refs.\ \cite{storzer06,sperling13} can be explained by weak fluorescent emission due to impurities present in TiO$_2$ powders without invoking Anderson localization \cite{sperling16}. They also convincingly demonstrate fluorescence and the resulting anomalous time-dependent behaviour 
in weakly disordered samples where localization can be excluded. They show that the fluorescence results in a red shift of the spectrum of scattered light with respect to the spectrum of the incident beam, and that the 
apparent signatures of localization are removed when this red-shifted light is filtered out of the detected signals. The authors conclude that the localization of light in a white paint, as proposed by Anderson \cite{anderson85}, is still unobserved and that the search for it should continue by getting rid of fluorescent impurities and optimizing the properties of the disorder.

Why has the search for Anderson localization of light in 3D been unsuccessful up to now, despite considerable efforts by several 
highly skilled, well-equipped and motivated groups? One of the reasons may be the limits imposed by the Nature on the values of the refractive indices of common transparent materials at optical frequencies: $n \lesssim 4$. This does not allow arbitrarily strong optical scattering to be achieved and can prevent reaching the Anderson localization transition. Another phenomenon playing against Anderson localization is the near-field coupling between scatterers that becomes important at high number densities of the latter \cite{skip14, rezvani15}. However, the failure of existing experiments does not mean that the phenomenon cannot be realized at all even though the ensemble of available results indicates that there is a need to go beyond fully disordered ``white paints''. A way to reach localization may consist in using partially ordered media such as photonic crystals with defects as proposed by Sajeev John \cite{john91}. Promising results were obtained following this proposal \cite{douglass11}. Another possibility would be to explore the potential of so-called hyperuniform structures that can be now fabricated routinely \cite{muller14}. One can also engineer scatterers with a complex internal structure (as opposed to grains of random shapes or spheres) that would have larger scattering cross-sections, one example being hollow spherical shells \cite{eiden04}. And finally the possibility of Anderson localization of light in a random ensemble of cold atoms under a strong magnetic field was theoretically predicted and awaits experimental realization \cite{skip15}.

After 30 years of research we have 
to conclude that Anderson localization of light in 3D still escapes experimental observation. In spite of this, the efforts of scientists working in this exciting field have impacted 
research on other types of classical waves such as, for instance, elastic waves for which the phenomenon has been observed \cite{hu08}. They also stimulated observation of Anderson transition for matter waves in cold \cite{chabe09} and ultra-cold \cite{jendr12} atomic systems. If observed, optical localization in 3D may have a great technological potential and may provide means to study the phenomenon with high accuracy and thus to confirm or to challenge the existing theoretical models.

\section*{Acknowledgements}

We thank the Agence Nationale de la Recherche (grant ANR-14-CE26-0032 LOVE), CNRS (France-Canada PICS project Ultra-ALT) and NSERC (Discovery Grant RGPIN/9037-2001) for financial support.

\end{document}